\documentclass[manuscript]{aastex}

\usepackage{color}
\usepackage[normalem]{ulem}

\shorttitle{The extinct $p$-nuclides: their origin from Type Ia supernovae and solar composition}
\shortauthors{Travaglio et al.}

\begin{document}


\title{Radiogenic $p$-isotopes from SN~Ia,
nuclear physics uncertainties and Galactic chemical evolution
compared with values in primitive meteorites}

\author{C. Travaglio\altaffilmark{1}}
\affil{INAF - Astrophysical Observatory Turin, Strada Osservatorio 20, 10025 Pino Torinese (Turin), Italy}
\affil{B2FH Association - Turin, Italy}
\email{travaglio@oato.inaf.it, claudia.travaglio@b2fh.org}

\author{R. Gallino\altaffilmark{2}}
\affil{Dipartimento di Fisica, Universit\'a di Torino, Via P.Giuria 1, 10125 Turin, Italy}
\affil{B2FH Association - Turin, Italy}

\author{T. Rauscher\altaffilmark{3}}
\affil{Centre for Astrophysics Research, School of Physics, Astronomy and Mathematics,
University of Hertfordshire, Hatfield AL10 9AB, United Kingdom}

\author{N. Dauphas\altaffilmark{4}}
\affil{Origins Laboratory, Department of the Geophysical Sciences and Enrico Fermi Institute, The University of Chicago, Chicago, IL 60637, USA}

\author{F. K. R{\"o}pke\altaffilmark{5}}
\affil{Universit{\"a}t W{\"u}rzburg, Am Hubland, D-97074 W{\"u}rzburg, Germany}

\and

\author{W. Hillebrandt\altaffilmark{6}}
\affil{Max-Planck-Institut f\"ur Astrophysik, Karl-Schwarzschild-Str.~1, D-85748 Garching bei M\"unchen, Germany}

\begin{abstract}

The nucleosynthesis of proton-rich isotopes is calculated for multi-dimensional
Chandrasekhar-mass models of Type Ia supernovae with different
metallicities. The predicted abundances of the short-lived radioactive isotopes
$^{92}$Nb, $^{97,98}$Tc, and $^{146}$Sm are given in this framework. The
abundance seeds are obtained by calculating $s$-process nucleosynthesis
in the material accreted onto a carbon-oxygen white dwarf
from a binary companion. A fine grid of $s$-seeds at
different metallicities and $^{13}$C-pocket efficiencies is considered. A
galactic chemical evolution model is used to predict the contribution of SN~Ia to the
solar system $p$-nuclei composition measured in meteorites. Nuclear physics
uncertainties are critical to determine the role of
SNe~Ia in the production of $^{92}$Nb and $^{146}$Sm. 
We find that, if standard Chandrasekhar-mass
SNe~Ia are at least 50\% of all SNIa, they are strong candidates for reproducing the radiogenic $p$-process
signature observed in meteorites.

\end{abstract}


\section{Introduction}

The astrophysical $p$-process is the conversion of a $s$- or $r$-process
distribution into proton-rich nuclei via photodisintegration reactions and
charged particle reactions.  This conversion can only occur on a
hydrodynamical timescale when the temperature is higher than 10$^9$ K.
Core-collapse supernovae (ccSN in what follows) and/or Type Ia supernovae
(SNe~Ia hereafter) are the most probable contributors to the bulk of the solar
system $p$-nuclei (for ccSN see, {\it e.g.}, Howard \& Meyer~1993, Rauscher et
al.~2002; for SN~Ia see {\it e.g.}, Travaglio et al.~2011 hereafter TRV11, Kusakabe et
al.~2011, Arnould \& Goriely~2006).

$p$-process nucleosynthesis occurs in SNIa by processing matter that was
enriched in $s$-process seeds during pre-explosive evolution of the SNIa
progenitor. Therefore, it is essential to determine the $s$-process enrichment
in the exploding white dwarf (WD hereafter). We consider a binary system,
accreting material from a giant star onto the WD. We explore the
single-degenerate scenario with a Chandrasekhar mass carbon-oxygen (CO-)
WD. The $s$-process seeds are assumed to be produced from a
sequence of thermal pulse instabilities in the accreted material. This idea
was described in detail by TRV11 and Kusakabe et al.~(2011), and previously
discussed by Iben~(1981), Iben \& Tutukov~(1991), Howard \& Meyer~(1993). Recurrent flashes are assumed to occur in the He-shell during the accretion
phase, resulting in enrichment of the CO-WD in $s$-nuclei . The mass involved in the $^{13}$C-pocket (a tiny layer
enriched in $^{13}$C responsible for the production of $s$-process nuclei) is
 a free parameter of the model. Since no model exists for the production of $s$-seeds in the accretion phase, 
different $s$-seed distributions are explored in order to better understand the dependence
of our results on the initial seed composition (see also the discussion in
TRV11).

The $p$-process produces radiogenic isotopes with relatively long half-lives
(Rauscher~2013). A number of now extinct short-lived nuclides were present in
the early solar system. Their past presence in meteorites is revealed by
measuring excesses of their decay-products in meteorites (Dauphas \&
Chaussidon~2011; Davis \& McKeegan~2013).

The isotope $^{92}$Nb decays with a half-life of 34.7 Myr to the stable nucleus $^{92}$Zr via $\beta$ decay. Harper~(1996) found first evidence for live
$^{92}$Nb in the early solar system material by measuring a small excess of
$^{92}$Zr in rutile (TiO\_2) extracted from the Toluca iron meteorite. Studies
of supernova neutrino nucleosynthesis (Hayakawa et al.~2013), alpha-rich
freezeout (Meyer~2003), or $\gamma$-process (Dauphas et al.~2003), tried to
explain the observed abundance of meteoritic $^{92}$Nb. Nevertheless, the
astrophysical site where the solar system $^{92}$Nb was made is still uncertain.

The isotope $^{146}$Sm decays to the stable isotope $^{142}$Nd by
$\alpha$-emission. Prinzhofer et al.~(1989, 1992) and Lugmair \& Galer~(1992)
provided the first solid estimates of the initial abundance of $^{146}$Sm at
the birth of the solar system.  The half-life of $^{146}$Sm is still under
debate.  The first measurements of its half-life was performed by
Friedmann~(1966) and later confirmed by Meissner et al.~(1987), setting a
value of $103\pm 5$ Myr. More recently Kinoshita et al.(2012), based on an
analysis of $^{146}$Sm/$^{147}$Sm $\alpha$-activity and atom ratios,
redetermined the half-life of $^{146}$Sm and found $68\pm 7$ Myr.

The isotopes $^{97}$Tc ($t_{1/2} = 4.21$ Myr) and $^{98}$Tc ($t_{1/2} = 4.2$
Myr) are of $p$-origin and may have been present in the early solar system 
but meteorite measurements only provide upper-limits on the abundances of these short-lived nuclides (Dauphas et
al.~2002; Becker \& Walker~2003).

In this work, we discuss the productions of $^{92}$Nb, $^{97,98}$Tc, and
$^{146}$Sm under SN~Ia conditions, their dependence on the astrophysical
environment, on the initial metallicity of the star, and on nuclear physics
quantities.

In Section \ref{sec:snIa}, the SN~Ia model and the method 
to compute nucleosynthesis in multi-D SN~Ia (TRV11 and references therein) are presented.  
In Section \ref{sec:calc} the $p$-process
calculations for radioactive isotopes and their dependence on
metallicity and $s$-seeds are discussed. In Section \ref{sec:gce} the galactic chemical evolution model
(Travaglio et al.~2004; Travaglio et al.~1999) together with calculations for radioactive
$p$-isotopes are presented.  Finally, in Section \ref{sec:nuclear} the sensitivity of
$^{146}$Sm and $^{92}$Nb SNIa yields to the uncertainties on rates as well as on their
lifetimes are discussed.

\section{Type Ia supernova nucleosynthesis and $p$-process radioactivities.}
\label{sec:snIa}

For the SN~Ia explosions, we use a delayed detonation model (DDT-a) based on
2-dimensional simulations of Kasen et al. (2009) and described in detail 
in TRV11. The scenario considered for SNIa is that of single-degenerate star, in which the WD accretes
material from a main-sequence or evolved companion star. Nucleosynthesis is
calculated in a post-processing scheme making use of tracer-particle
methods (as described in detail in TRV11 and Travaglio et al.~2004b). For each
tracer, explosive nucleosynthesis is followed using a detailed nuclear
reaction network for all isotopes up to $^{209}$Bi. The nuclear
reaction rates used are based on the experimental values and the Hauser \& Feshbach
statistical model NON-SMOKER (Rauscher \& Thielemann~2000), including the
experimental results of Maxwellian averaged neutron-capture cross sections of
various $p$-only isotopes (Dillmann et al.~2010; Marganiec et
al~2010). Theoretical and experimental electron capture and $\beta$-decay
rates are from Langanke \& Mart\'inez-Pinedo~(2000).

TRV11 demonstrated that the
abundances of the $p$-nuclei in SNIa strongly depend on the $s$-seeds assumed.
However, the ratio of a radiogenic isotope to the neighbor stable $p$-isotope
(i.e. $^{92}$Nb/$^{92}$Mo, $^{97,98}$Tc/$^{98}$Ru and $^{146}$Sm/$^{144}$Sm),
is less dependent of the assumptions made for the $s$-seeds. Note that all
the reference stable isotopes are pure $p$-nuclei also.

\begin{figure}
\includegraphics[width=\columnwidth]{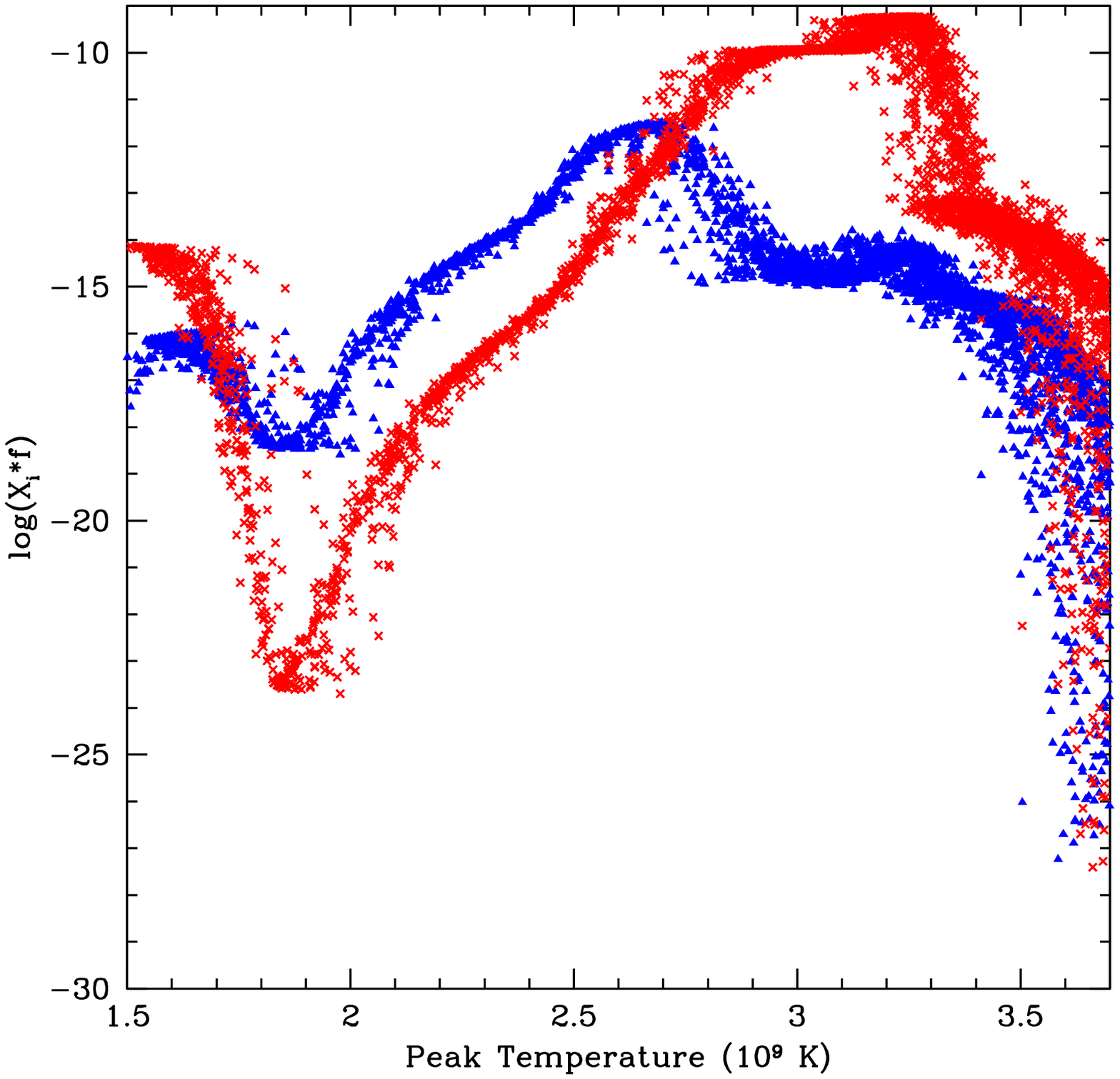}
\caption{\label{fig:abunNbMo}Abundance vs. $T$ peak of the ratio $^{92}$Nb and $^{92}$Mo for tracers selected in the $T$ range allowed for $p$-process nucleosynthesis. Each small dot
represents one tracer. {\it Filled blue triangles} are for $^{92}$Nb and {\it red crosses} are for $^{92}$Mo. All the abundances $X_i$ shown here and in the following
Figures~2 and 3 are for an individual tracer, and the f factor in the plot is for $M_\mathrm{WD}$(= 1.407 M$_\odot$)/N$_{tracers}$ (=51200), i.e. the mass of each tracer.} \end{figure}

\begin{figure}
\includegraphics[width=\columnwidth]{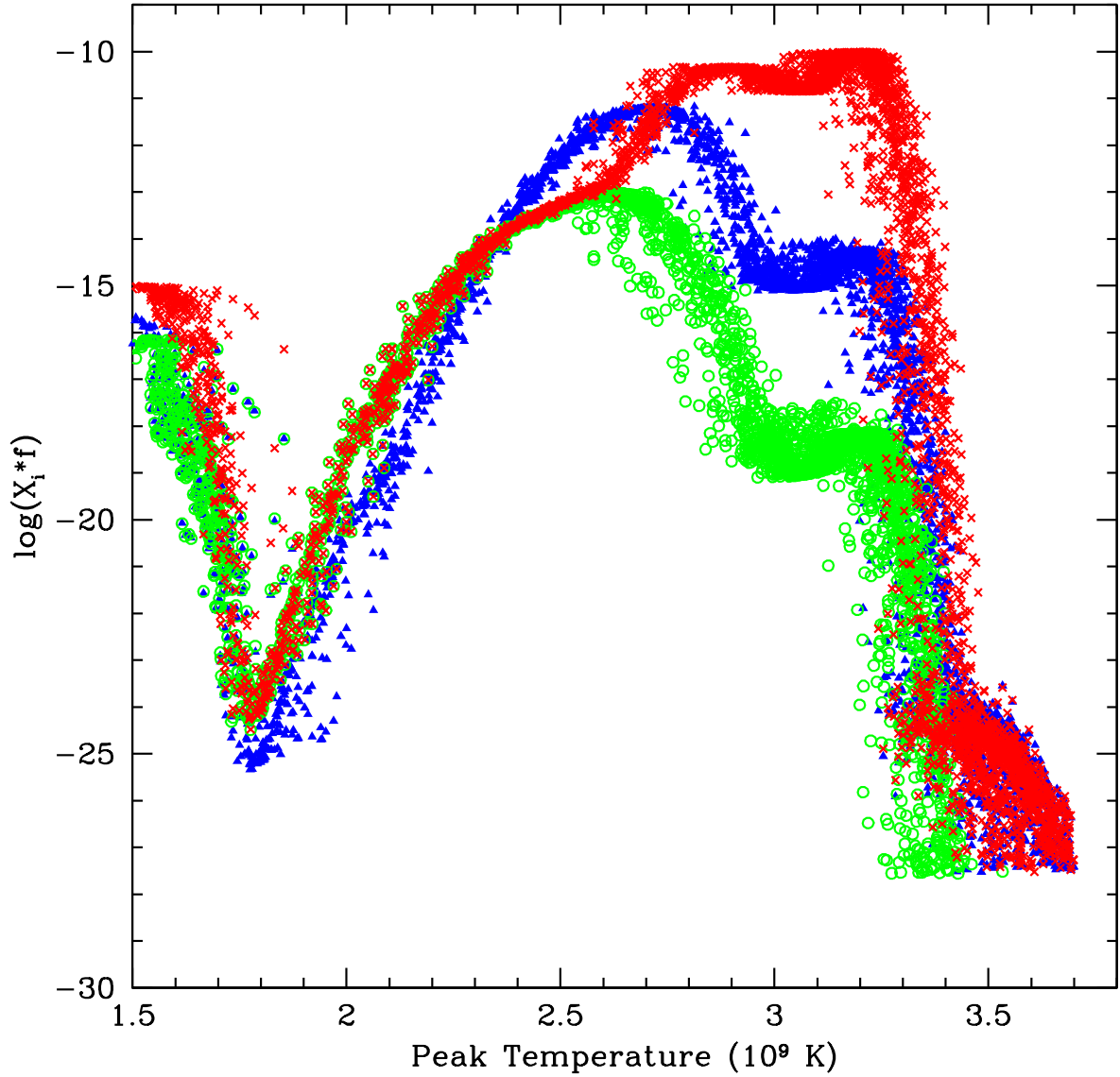}
\caption{\label{fig:abunTcRu}Same as Fig.\ \ref{fig:abunNbMo}, for $^{97}$Tc ({\it filled blue triangles}), $^{98}$Tc ({\it open green circles}) and $^{98}$Ru ({\it red crosses})} 
\end{figure}

\begin{figure}
\includegraphics[width=\columnwidth]{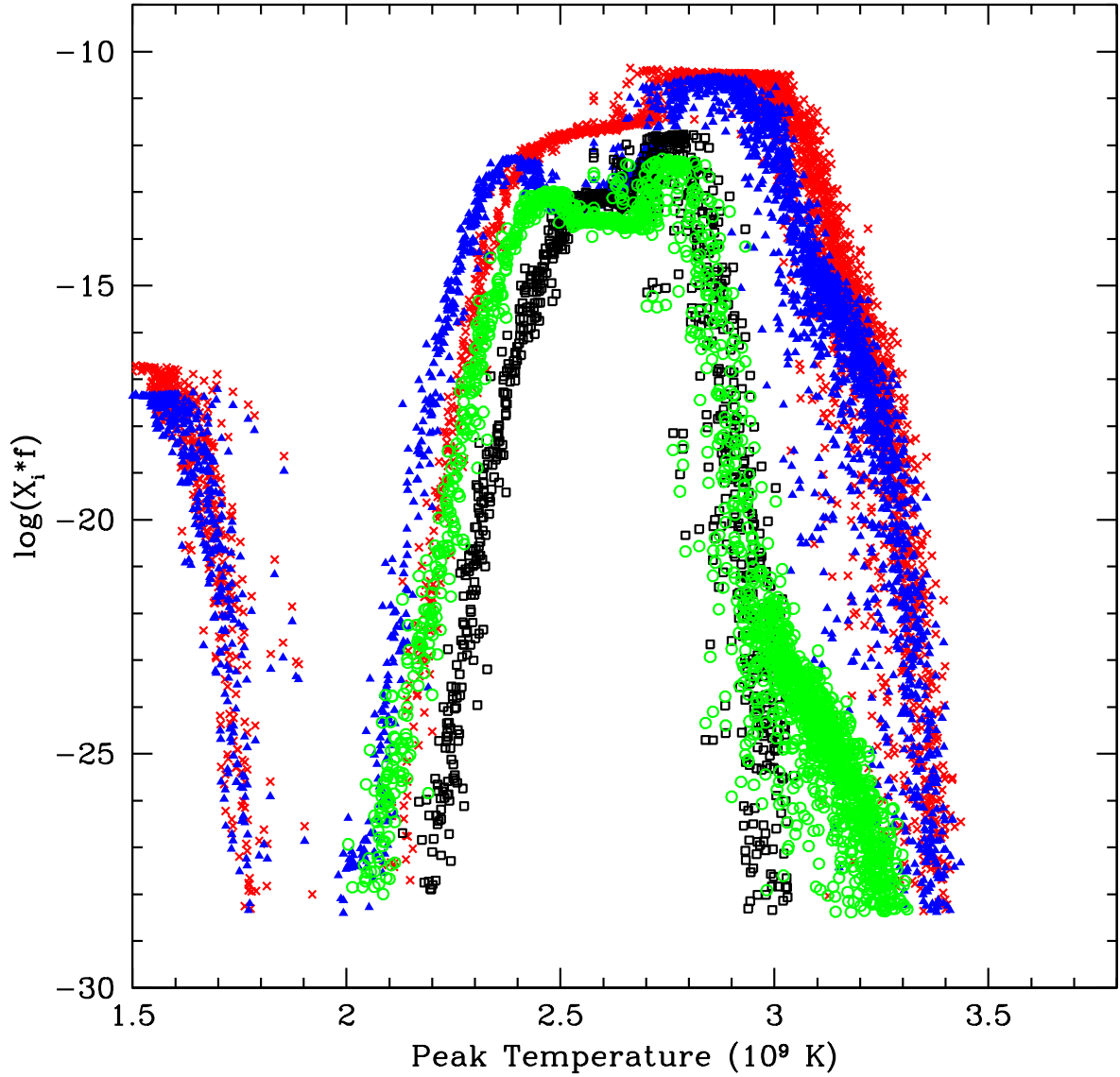}
\caption{\label{fig:abunSmGdDy}Same as Fig.\ \ref{fig:abunNbMo}, for $^{146}$Sm ({\it filled blue triangles}), $^{150}$Gd ({\it open green circles}), $^{154}$Dy ({\it open black squares}) and 
$^{144}$Sm ({\it red crosses})} \end{figure}

The abundances of $^{92}$Nb, $^{92}$Mo, $^{97,98}$Tc,$^{98}$Ru and
$^{146}$Sm,$^{144}$Sm obtained in this way are plotted in
Figs.\ \ref{fig:abunNbMo}$-$\ref{fig:abunSmGdDy} for tracers selected in the
peak temperature range that allows $p$-process nucleosynthesis ({\it i.e.}, $1.5 -
3.7$ GK).  Each dot represents one tracer at its peak
temperature. In the DDT-a model, we have 51200 tracer particles in total, and
4624 of them in the $p$-process temperature range located in the accreted
mass, representing a total mass of 0.127 $M_\odot$ (the mass of a single tracer is
$2.75 \times 10^{-5}$ $M_\odot$). Since we verified that an important
contribution to $^{146}$Sm comes from the decay of $^{150}$Gd and of
$^{154}$Dy, we did not include them directly in the $^{146}$Sm abundance
plotted but they are shown separately in Fig.\ \ref{fig:abunSmGdDy}.
As can be seen in Fig.\ \ref{fig:abunNbMo}, most of the production of $^{92}$Nb
takes place at around $T = 2.5 - 2.7$ GK, where $^{20}$Ne burning occurs (see Figure~5 in
TRV11 for details).

$s$-process distributions are calculated for a fine
grid of metallicities, {\it i.e.}, $Z =0.02, 0.015, 0.012, 0.010, 0.006, 0.003$, and
for different $^{13}$C-pockets (ST$\times$2, ST$\times$1.3, ST, ST/1.5, where
$\sim4 \times 10^{-6}$ $M_\odot$ of $^{13}$C in the pocket corresponds to the
ST case, Gallino et al.~1998).  The mass involved and the profile of the
$^{13}$C mass fraction are treated as free parameters. In the
present state of the art, there is no model that can predict $s$-process
nucleosynthesis during the mass-accretion phase prior a CO-WD explosion. TRV11
presented and discussed different possible $s$-process seed distributions in
mass-accretion conditions, and their consequences for $p$-process
nucleosynthesis. Here and in Travaglio et al.~(2014, in prep.), we add
a detailed analysis of the dependence on metallicity.

The goal of the present work is to provide predictions for solar composition
of radioactive $p$-nuclei. The Galactic chemical evolution code used has been
presented already in previous studies (Travaglio et al.~1999, 2001, 2004,
Bisterzo et al.~2014), see Section~4 for a detailed discussion.

\section{$p$-process and radioactive $^{92}$Nb, $^{97,98}$Tc, and $^{146}$Sm
for different metallicities}
\label{sec:calc}

Extinct radionuclides were found in meteorites (Dauphas \& Chaussidon~2011;
Davis \& McKeegan~2013) and some of them are $p$-only radiogenic nuclei,
{\it i.e.}, $^{92}$Nb and $^{146}$Sm. Their signatures were detected as an excess
abundance of the daughter nuclei ($^{92}$Zr and $^{146}$Nd).

Whereas $^{93}$Nb is 85\% $s$-process and 15\% $r$-process (Arlandini et al.~1999),
$^{92}$Nb is an important isotope since it is produced by the $\gamma$-process
but is completely shielded from contributions from $rp$- or ${\nu}p$-processes
(Dauphas et al.~2003), and as such can help test models of $p$-process
nucleosynthesis. Meteorite measurements show that this nuclide was present at
the birth of the solar system (with an initial $^{92}$Nb/$^{92}$Mo ratio of
(2.80$\pm$0.5)$\times$10$^{-5}$ (Harper~1996; Sch\"onb\"achler et al.~2002;
Rauscher et al.~2013). Nevertheless its astrophysical production site is still
unknown (Dauphas et al.~2003; Meyer~2003). Note that $^{92}$Nb is normalized
to $^{92}$Mo because both are $p$-process nuclides while $^{93}$Nb is mainly a
$s$-process isotope (by the radiogenic decay of $^{93}$Zr).

The underproduction of $^{92,94}$Mo and $^{96,98}$Ru in the $\gamma$-process
could, in principle, be compensated by contributions of the $rp$- or
${\nu}p$-processes but this would lead to a too low $^{92}$Nb/$^{92}$Mo ratio
at solar system birth (Dauphas et al.~2003). Various theoretical estimates for
the ratio $^{92}$Nb/$^{92}$Mo are available in literature, for both SN~Ia
(Howard et al.~1991; Howard \& Meyer~1993) and core-collapse supernovae (ccSN)
(Woosley \& Howard~1978; Woosley \& Howard~1990; Rayet et al.~1995; Hoffman et
al.~1996; Rauscher et al.~2002; Hayakawa et al.~2013). Some of these models
can reproduce the solar system $^{92}$Nb/$^{92}$Mo ratio but at the same time,
they underproduce the amount of $^{92}$Mo present in cosmic abundances, so
they are unlikely contributors to $p$-process nuclides in the Mo-Ru mass
region (Rauscher et al.~2002). 

The short-lived $^{146}$Sm was also present in meteorites at the birth of the solar
system. The first attempts to estimate the $^{146}$Sm/$^{144}$Sm ratio were
published by Lugmair \& Marti~(1977) and by Jacobsen \& Wasserburg~(1984). The
initial $^{146}$Sm/$^{144}$Sm ratio was constrained in meteorites by Lugmair
\& Galer~(1992) and by Prinzhofer et al.~(1992). A later study of eucrite
meteorites by Boyet et al.~(2010) confirmed the value found but improved the
precision, contraining the $^{146}$Sm/$^{144}$Sm ratio at the birth of the
solar system to 0.0084 $\pm$ 0.0005 using a half-life of 103 Myr to correct
for decay between formation of the eucrites and the solar system. Using the
half-life for $^{146}$Sm of 68 Myr and the last meteorite measurements,
Kinoshita et al. (2012) estimated that the initial $^{146}$Sm/$^{144}$Sm ratio
was 9.4 $\pm$ 0.5 $\times$ 10$^{-3}$.  Because eucrite crystallization
occurred shortly after the formation of the solar system, the correction of
the initial $^{146}$Sm/$^{144}$Sm ratio is small and does not depend much on
which $^{146}$Sm half-life is used.

The nuclides $^{97,98}$Tc have not been detected in meteorites yet, and only
upper limits were derived. Dauphas et al.~(2002) provided an upper-limit on the
$^{97}$Tc/$^{98}$Ru ratio of $<$ 4$\times$ 10$^{-4}$. Becker \& Walker~(2003) provided an upper-limit on the $^{98}$Tc/$^{98}$Ru ratio of $<$ 2$\times$ 10$^{-5}$.

In Table \ref{tab:ratios} we show the values of the radiogenic ratios
$^{92}$Nb/$^{92}$Mo, $^{97,98}$Tc/$^{98}$Ru, and $^{146}$Sm/$^{144}$Sm
obtained with the tracer-particle nucleosynthesis calculations based on the DDT-a
model as a function of metallicity. $s$-process distributions are calculated for a range of galactic disk
metallicities, from solar value down to $Z$ = 0.003. The
effect of using different $^{13}$C-pocket sizes is also explored (ST$\times$2, ST$\times$1.3,
ST, ST/1.5, where $\sim$4 $\times$ 10$^{-6} M_\odot$ of $^{13}$C in the pocket
corresponds to the ST case, Gallino et al.~1998). The parameter used for Table
\ref{tab:ratios} is ST$\times$1.3. The GCE calculations and comparisons with meteoritics abundances are described in details in the following section.

\section{Galactic chemical evolution}
\label{sec:gce}

The initial $^{92}$Nb/$^{92}$Mo and
$^{146}$Sm/$^{144}$Sm ratios at the birth of the solar system are known from meteorite measurements. In order to compare
these values with our model results, one has to use a model of chemical evolution of the Galaxy, as the
abundances in the ISM at any given time reflect the interplay between
production in stars and decay in the ISM. Note that we do not consider any contribution from stars other than SNIa to the nucleosynthesis of $p$-nuclides.

The Galactic chemical evolution code used here was presented in several
publications (Travaglio et al.~1999; Travaglio et al.~2001; Travaglio et al.~2004;
Bisterzo et al.~2014). It models the Galaxy as 
three interconnected zones; halo, thick disk, and thin disk. The evolution of the Galaxy 
is computed up to the present epoch ($t_\mathrm{today} =
13.8$ Gyr, updated by WMAP, Bennet et al.~2013) and the solar system formation
is assumed to have occurred 4.6 Gyr ago. Therefore the time
corresponding to the birth of the solar system is $t_\odot = 9.2$ Gyr. Solar
abundances are taken from Lodders et al.~(2009) and massive star yields are from Rauscher et
al.~(2002). Iron is mostly produced by long-lived SNe~Ia (a knee in the trend
of [O/Fe] vs. [Fe/H] indicates the delayed contribution to Fe by
SNe~Ia, see e.g. McWilliam~1997). Following the common idea that oxygen is mainly synthesized by
short-lived massive stars in ccSN and Fe is mostly produced by long-lived
binary systems in the form of SNe~Ia, the knee in the observed trend of [O/Fe]
vs [Fe/H] (in field stars at different metallicities) indicates the delayed
contribution to iron by SNe~Ia (1/3 of Fe is probably produced by ccSN and 2/3
by SN~Ia). As recently shown by Bisterzo et al.~(2014), with an updated
compilation of spectroscopic data, our model fits well the knee observed,
giving a good constraint to the rate of ccSN {\it vs}. SN~Ia, as well as to the
treatment of binary stars included in the GCE code (for details see Travaglio
{\it et al.}~1999).

The matrix of isotopes within the chemical evolution code was set to cover
all the light nuclei up to the Fe-group, and all the heavy nuclei along the
$s$-process and $p$-process paths up to $^{209}$Bi. The resulting $p$-process
production factors taken at the epoch of solar system formation for nuclei in
the atomic mass number range 70 $\le A \le$ 210 are shown in detail in
Travaglio et al.~(2014, in prep.), with a fine grid of metallicities and exploring different
$s$-process seed distributions. In this paper, the choice for $s$-seed distribution 
versus metallicity is discussed 
in detail, {\it i.e.}, we choose higher $^{13}$C
for higher metallicities (ST$\times$2 for solar metallicity, ST$\times$1.3 for
metallicities down to 0.01, ST for $Z =$ 0.006, and ST/1.5 for the lowest
metallicities). The grid of metallicities used for the present work is described in
Section \ref{sec:calc} (see also Table \ref{tab:ratios}). With this choice,
the predicted ratios at 9.2 Gyr are
$^{92}\mathrm{Nb}/^{92}\mathrm{Mo} = 1.752 \times 10^{-5}$ (about a factor of
1.6 below the meteoritic value of $2.8 \pm 0.5 \times 10^{-5}$),
$^{146}\mathrm{Sm}/^{144}\mathrm{Sm} = 6.989 \times 10^{-3}$ (about a factor
of 1.3 below the meteoritic value of $9.4 \pm 0.5 \times 10^{-3}$),
$^{97}\mathrm{Tc}/^{98}\mathrm{Ru} = 4.077 \times 10^{-5}$, and
$^{98}\mathrm{Tc}/^{98}\mathrm{Ru} = 6.471 \times 10^{-7}$ (for Tc the ratios
measured in meteorites are upper limits). The ratios as a function of the age
of the Galaxy (or of the metallicity) are given in Table \ref{tab:gce}. The meteoritic values measured and their
errors are
also reported in the same table . A detailed analysis and discussion on the importance of uncertainties
of reaction rates for $^{92}$Nb and $^{146}$Sm is presented in the next section.

Most previous studies dealing with radioactive nuclides in meteorites have relied on analytical or semi-analytical approaches to predict the abundances of 
these nuclides in the ISM at solar system birth (Schramm \& Wasserburg~1970; Clayton~1988; Dauphas et al.~2003; Dauphas~2005; Huss et al.~2009; 
Jacobsen~2005). The main virtue of these analytical approaches is that they allow one to rapidly explore the parameter space while capturing some of the 
most important features of galactic chemical evolution. However, this simplicity is achieved at the expense of realisticness. Analytical approaches can 
take into account the secondary nature of some nuclides, the fact that the galactic disk probably grew by infall of low-metallicity gas, and the fact that 
the rate of star formation is not strictly linear with the gas surface density. However, all analytical models rely on the instantaneous recycling 
approximation, which assumes that material newly produced in stars is immediately returned to the ISM. This assumption is not correct for nucleosynthesis 
in SNIa, and would produce a factor of 2 to 4 higher values of the radiogenic ratios discussed above, but also wrong predictions of Fe at solar composition.
More sophisticated galactic chemical models such as that presented here should be used. 

\section{Nuclear and half-life uncertainties}
\label{sec:nuclear}

\begin{figure}
\includegraphics[width=\textwidth]{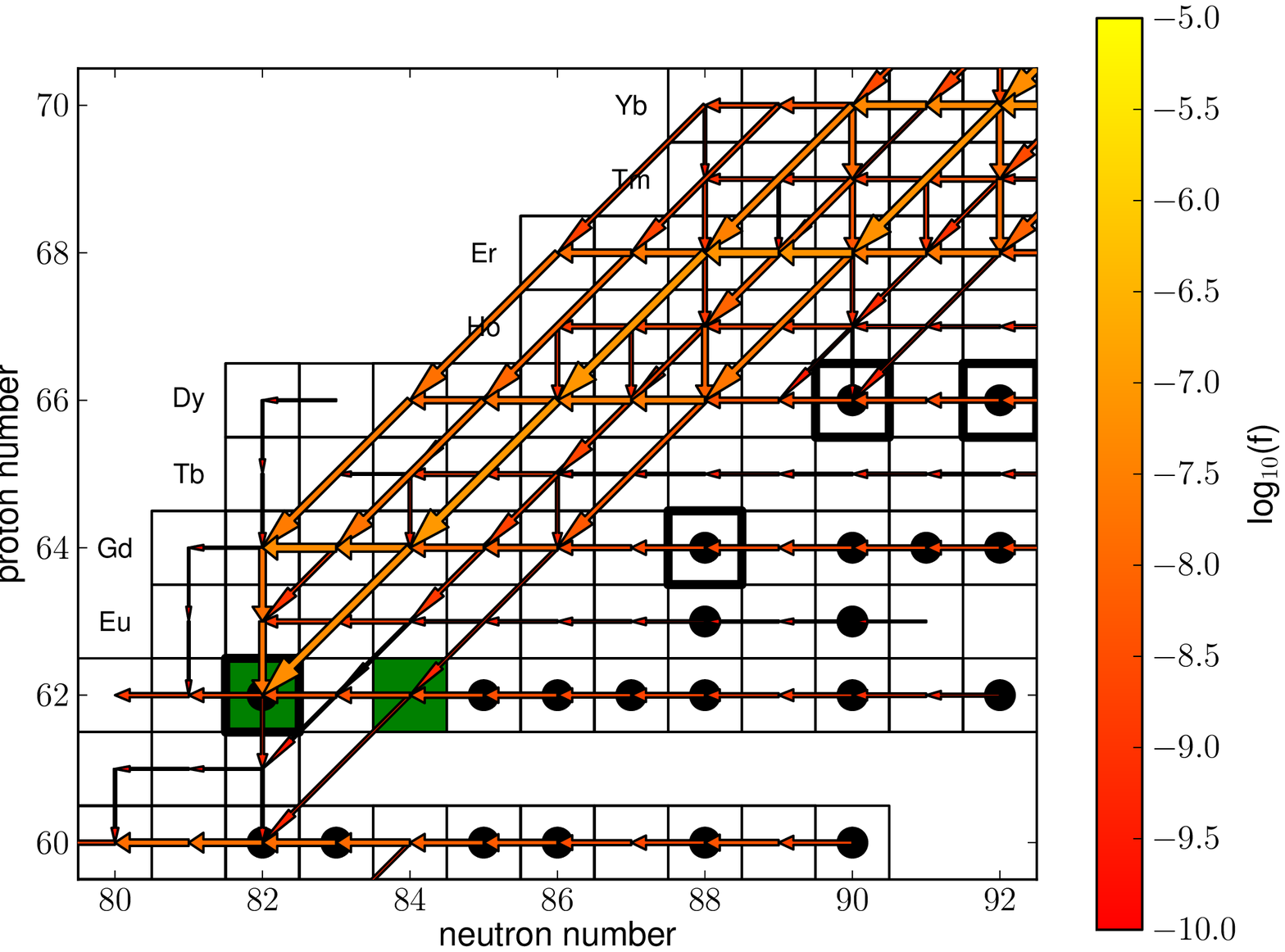}
\caption{\label{fig:flowSm}Reaction flow for $^{146}$Sm production; size and color of the arrows relate to the magnitude of the time-integrated flux on a logarithmic scale. 
Stable isotopes are marked by a {\it black dot} and $p$-isotopes by a {\it thicker box}. The isotopes $^{144}$Sm and $^{146}$Sm are marked in {\it green}.}
\end{figure}

Figure \ref{fig:flowSm} shows the reaction flow (\textit{i.e.},
time-integrated flux) for a selected tracer producing 
$^{146}$Sm and $^{144}$Sm isotopes.  As can be seen in the figure, the main flow from heavier nuclei is
following the line defined by mass number $A=Z+82$, with $Z$ being the nuclear
charge. The production ratio of $^{146}$Sm/$^{144}$Sm mainly depends on the ($\gamma$,n)/($\gamma$,$\alpha$)
branching at $^{148}$Gd (and more weakly on similar branchings at $^{152}$Dy and
$^{156}$Er). This is due to the fact that $^{146}$Gd is neutron magic and, after the passage of the shock wave,
decays to $^{146}$Eu and then to $^{146}$Sm. This was already
pointed out by Woosley \& Howard~(1990) and further investigated by Rauscher
et al.~(1995) and Somorjai et al.~(1998) for the $\gamma$-process in massive
stars.  This branching is independent of the seeds but a weak dependence of
the $^{146}$Sm/$^{144}$Sm ratio stems from the production of $^{144}$Sm
and the weak ($\gamma$,n) flow in Sm. This weak dependence is seen in Table
\ref{tab:gce}, where the $^{146}$Sm/$^{144}$Sm ratio obtained for different
metallicities is shown and also in Table \ref{tab:sm} where additionally the
dependence on various $^{148}$Gd($\gamma$,$\alpha$)$^{144}$Sm rates is
presented (also see discussion below).

Photodisintegration rates at high plasma temperature cannot be constrained by
direct measurements (Rauscher~2012; Rauscher~2014). A better test of predicted
reaction cross sections and astrophysical reaction rates could be obtained by
experimentally determining capture cross sections. Since $^{147}$Gd, however,
is an unstable nucleus with a half-life of 38.06 hr, a measurement of
$^{147}$Gd(n,$\gamma$)$^{148}$Gd is not feasible and thus its rate has to be
derived from Hauser-Feshbach theory. For the calculations shown here, the rate by Rauscher \& Thielemann~(2000) was used. Comparison to neutron capture
data along the valley of beta-stability showed that the averaged uncertainty of the predictions
was about 30\% but local deviations up to a factor of 2 were possible
(Rauscher et al.~1997; Rauscher et al.~2001; Rauscher~2012).

The measurement of the low-energy ($\alpha$,$\gamma$) cross section of the
stable $^{144}$Sm nucleus by Somorjai et al.~(1998) sets
the stage for a long-standing puzzle regarding the prediction of low-energy
$\alpha$-capture and emission. The cross section was found to be lower by more
than an order of magnitude than all predictions. Over the past years, many
attempts have been made to construct improved $\alpha$+nucleus optical
potentials to explain these data (see, {\it e.g.}, Kiss {\it et al}.~2013; and Rauscher {\it et
al}.~2013) and references therein. Recently, Rauscher~(2013) suggested that
including an additional reaction channel, only affecting $\alpha$-capture but
not emission, the experimental results can be reproduced. In Table \ref{tab:sm}, the $^{146}$Sm/$^{144}$Sm ratios obtained
with three different $^{148}$Gd($\gamma$,$\alpha$)$^{144}$Sm rate predictions
are shown: the original rate by Rauscher \& Thielemann~(2000) based on a cross section higher
than the experimental value, the rate based on a fit to the Somorjai et
al.~(1998) ($\alpha$,$\gamma$) cross sections, and the new rate by
Rauscher~(2013) which reproduces the measured cross sections but predicts a
larger $\alpha$-emission rate. In ccSN, the Rauscher \& Thielemann~(2000) rate gave a too low
$^{146}$Sm/$^{144}$Sm ratio compared to meteoritic values, the fit to Somorjai
et al.~(1998) gave a much larger ratio, while the new rate again provided a
lower ratio due to the stronger $\alpha$-emission (Rauscher~2013; Rauscher et
al.~2013).  All of these values were just barely compatible or incompatible
with meteoritic ratios. For the SNIa case studied here, the final
$^{146}$Sm/$^{144}$Sm ratio integrated over the GCE is found to be compatible
when using the rate fit by Somorjai et al.~(1998) but also the new rate by
Rauscher~(2013). The difference with respect to the ccSN results is due to the
different temperature history of the tracer in SN~Ia as compared to the
situation in a ccSN shock front, since the ($\gamma$,n)/($\gamma$,$\alpha$)
branching at $^{148}$Gd is temperature dependent. It should be noted that
these ratios still bear an additional uncertainty from the
$^{148}$Gd($\gamma$,n) rate, as discussed above.

All the calculations presented here used the most recent $^{146}$Sm half-life
of 68 Myr (Kinoshita et al.~2012). However, it is worth noting that the value
of this half-life is still the subject of ongoing discussions as comparisons
of $^{146}$Sm-$^{142}$Nd with Pb-Pb or $^{147}$Sm-$^{143}$Nd dating techniques
may be more consistent with a longer half-life of 103 Myr (Borg et
al.~2014). This half-life is important in early solar system chronology and it
should be remeasured to ascertain its value.

\begin{figure}
\includegraphics[width=\textwidth]{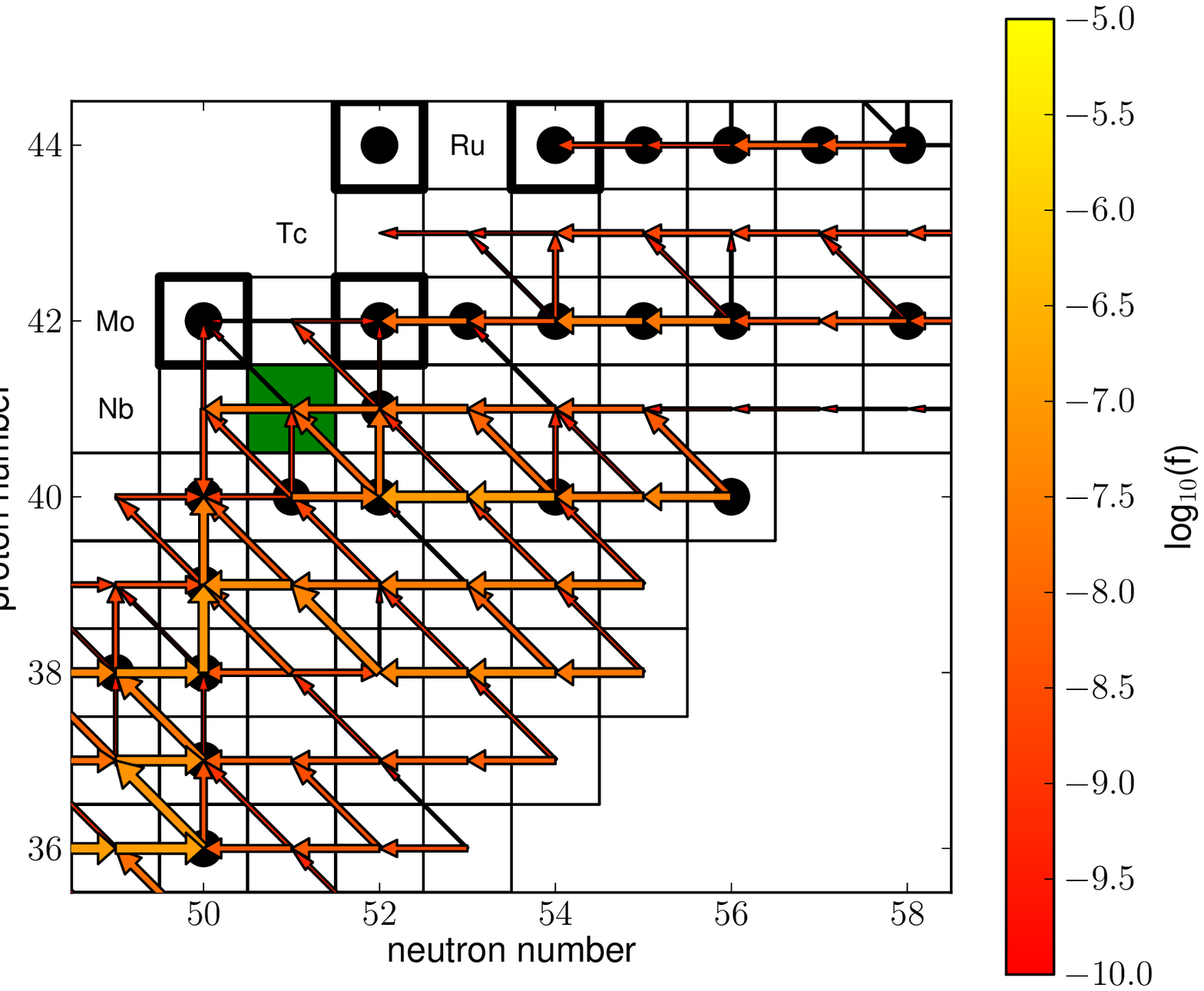}
\caption{\label{fig:flowNb}Reaction flow for $^{92}$Nb production; size and color of the arrows relate to the magnitude of the time-integrated flux on a logarithmic scale. Stable isotopes
are marked by a {\it black dot} and $p$-isotopes by a {\it thicker box}. The nuclide $^{92}$Nd is marked in {\it green}.}
\end{figure}

The production of the radiogenic $^{92}$Nb is governed by the destruction of
$^{93}$Nb and $^{92}$Zr seeds, as can be seen from the flows in
Fig.\ \ref{fig:flowNb}. It also gets some indirect contributions from
$^{91,94,96}$Zr via $^{92}$Zr but none from $^{90}$Zr. The nuclide $^{92}$Nb
is mainly destroyed by the reaction $^{92}$Nb($\gamma$,n)$^{91}$Nb, while two
reactions produce it, $^{93}$Nb($\gamma$,n)$^{92}$Nb and
$^{92}$Zr(p,n)$^{92}$Nb. A minor production channel (about 3\%) is
$^{91}$Zr(p,$\gamma$)$^{92}$Nb. Because the two reactions destroying $^{92}$Zr --
$^{92}$Zr(p,n) and $^{92}$Zr(p,$\gamma$) -- both eventually lead to $^{92}$Nb
production, their relative magnitude is not important, only their combination
into a total rate. The production of $^{92}$Zr proceeds via ($\gamma$,n)
sequences from the other Zr isotopes.  The slowest reactions in these
sequences are the ones removing a paired neutron and thus they dominate the
timescale and the flow. Here, this is $^{94}$Zr($\gamma$,n)$^{93}$Zr and, with slightly
less importance, $^{96}$Zr($\gamma$,n)$^{95}$Zr, both leading to eventual production of
$^{92}$Zr. Finally, $^{94}$Nb($\gamma$,n)$^{93}$Nb is important in the
production of $^{93}$Nb from neutron-richer Nb isotopes.

The rates of $^{94}$Zr($\gamma$,n)$^{93}$Zr and
$^{94}$Nb($\gamma$,n)$^{93}$Nb are experimentally determined through their
measured neutron capture cross sections (Kadonis, Dillmann et
al.~2006). Despite of the elevated temperatures found in $\gamma$-process
nucleosynthesis, the experimental data constrains both capture and
photodisintegration well in these cases (Rauscher~2012). For the other rates
given above, and their reverse reactions, we used predictions by Rauscher \& Thielemann~(2000)
in our standard calculations. The $^{96}$Zr($\gamma$,n)$^{95}$Zr rate comes from a
theory estimate as given in KADoNiS (Dillmann et al.~2006, Bao et al.~2000).

\begin{figure}
\includegraphics[width=\textwidth]{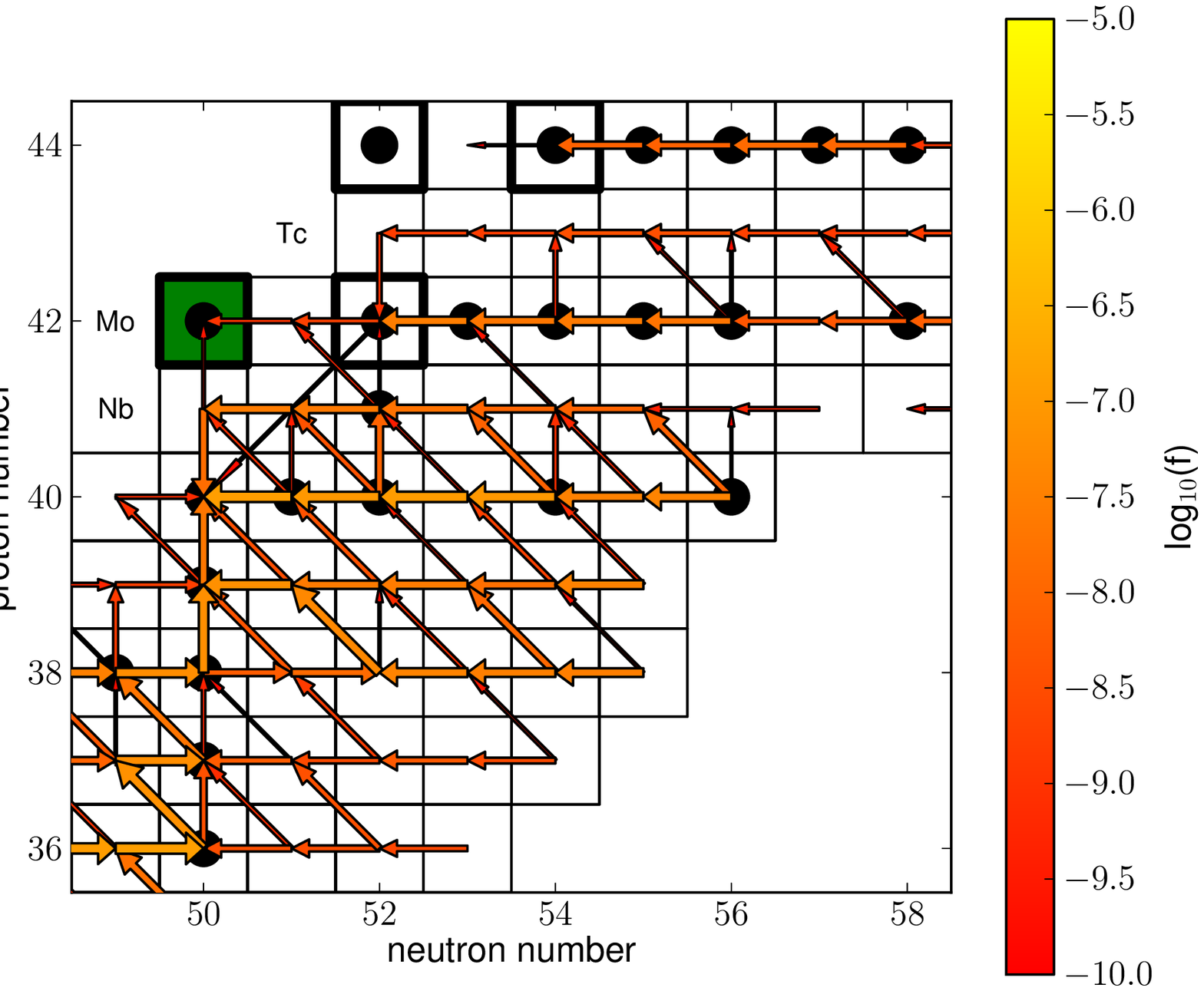}
\caption{\label{fig:flowMo}Reaction flow for $^{92}$Mo production; size and color of the arrows relate to the magnitude of the time-integrated flux on a logarithmic scale. Stable isotopes
are marked by a {\it black dot} and $p$-isotopes by a {\it thicker box}. The nuclide $^{92}$Mo is marked in {\it green}.}
\end{figure}

The uncertainty in the $^{92}$Nb/$^{92}$Mo ratio also contains the uncertainty
in the $^{92}$Mo production. Figure \ref{fig:flowMo} shows the time-integrated
flows in the tracer that produces the main fraction of $^{92}$Mo. The flow
pattern is less complex than in the case of $^{92}$Nb. The main contribution
to this nuclide (about 50\%) is through ($\gamma$,n) sequences coming from the
stable Mo isotopes with mass numbers $A>94$. These are mainly producing
$^{94}$Mo, part of which is converted to $^{92}$Mo through the reaction
sequence $^{94}$Mo($\gamma$,n)$^{93}$Mo($\gamma$,n)$^{92}$Mo. The slower
reaction in this sequence, determining the flow is $^{94}$Mo($\gamma$,n)$^{93}$Mo,
leaving an unpaired neutron in $^{93}$Mo. The second important path,
contributing about 35\%, is the sequence
$^{93}$Nb(p,n)$^{93}$Mo($\gamma$,n)$^{92}$Mo. Although the magnitude of the
(p,n) reaction also scales with the proton density, the $^{93}$Mo($\gamma$,n)$^{92}$Mo
reaction is the faster one again in this sequence at our SNIa conditions.
Finally, the reaction $^{91}$Nb(p,$\gamma$) provides a small (15\%),
additional contribution to $^{92}$Mo. There are only theoretical predictions
for the rates that are important, $^{94}$Mo($\gamma$,n)$^{93}$Mo, $^{93}$Nb(p,n)$^{92}$Mo, and
(with lower impact) $^{91}$Nb(p,$\gamma$)$^{92}$Mo. The $^{92}$Mo production would
scale according to the above weights when new rate determinations become
available for these reactions. 

The important theoretically estimated rates affecting the production of
$^{92}$Nd and $^{92}$Mo are summarized in Table \ref{tab:nbmoratio}. In order
to check the uncertainty in our GCE calculations at the solar system birth for
the $^{92}$Nb/$^{92}$Mo ratio due to uncertainties in the reaction rates, we
varied the most important rates found above by a factor of two. Calculations
with two rate sets were performed, probing the extremal values expected for
the $^{92}$Nb/$^{92}$Mo ratio as indicated in Table \ref{tab:nbmoratio}. This
leads to a $^{92}$Nb/$^{92}$Mo ratio at 9.2 Gyr varying from 1.660$\times$
10$^{-5}$ for the {\it Rate set MIN} up to 3.118$\times$ 10$^{-5}$ for the
{\it Rate set MAX} (the results are summarized in the last line of Table \ref{tab:nbmoratio}).

As shown in Table 2, the SNIa yields calculated here, when folded in a galactic chemical evolution, given nuclear physics uncertainties, we can reproduce 
the $^{92}$Nb/$^{92}$Mo and $^{146}$Sm/$^{144}$Sm ratios at solar system birth measured in meteorites ($^{92}{\rm Nb}/^{92}{\rm Mo}=2.8\times 10^{-5}$ in 
meteorites {\it vs}. 1.8$\times 10^{-5}$ predicted; ($^{146}{\rm Sm}/^{144}{\rm Sm}=9.4\times 10^{-3}$ in meteorites {\it vs}. 7.0$\times 10^{-3}$ 
predicted). 
Note that the match between predicted and measured ratios requires that the material that made the solar system had not been isolated from fresh 
nucleosynthetic inputs for some extended time, as is observed for some $r$-process short-lived nuclides such as $^{129}$I (Qian et al. 1998). These authors 
concluded that the discrepancy can be solved if some $r$-process isotopes are produced in rare events only. This study supports the view that 
single-degenerate SNIa may be important contributors to the nucleosynthesis of $p$-process nuclides in the Galaxy.

\section{Conclusions}

In this work we discuss the production of short-lived radionuclides $^{92}$Nb,
$^{146}$Sm and $^{97,98}$Tc by single degenerate SNIa stars. Using a simple Galactic chemical
evolution code, we show that a significant fraction of $p$-process extinct radionuclides $^{92}$Nb, $^{146}$Sm, and $^{96,98}$Tc in
meteorites could have been produced by the $\gamma$-process in SNeIa.

Travaglio et al. (2011) showed that SNIa were likely sites for $p$-process
nucleosynthesis. In particular, enrichment in $s$-seeds during the
pre-explosive evolution leads to the production of $^{92}$Mo, $^{94}$Mo, $^{96}$Ru
and $^{98}$Ru, $p$-process isotopes that exist in high abundance in the cosmos
and are difficult to reproduce in previous nucleosynthesis models. Dauphas et
al. (2003) pointed out that a critical test that models of $p$-process
nucleosynthesis must pass is that they must reproduce the abundances of the
short-lived nuclides $^{92}$Nb and $^{146}$Sm. In particular, $^{92}$Nb 
provides strong constraints on $p$-process nucleosynthesis because 
it is shielded by $^{92}$Mo from decays of proton-rich progenitor nuclides and thus cannot be produced
by processes on the proton-rich side of the nuclear chart, such as the $rp$- or $\nu p$-processes.

We should note that the nature of SN~Ia progenitors remains still uncertain. Following 
the idea of Li et al.~(2011) and Jimenez et al.~(2013), we supposed that at least 50\% of SN~Ia are single degenerate standard Chandrasekhar mass. 
But the reader has to keep in mind that they can be more rare. Referring to Ruiter et al.~(2013) and Ruiter et al.~(2014), 
population synthesis models tells us that SN~Ia progenitors come from a (rare) sample of common-envelope phase binaries which may or may not undergo some 
$s$-processing before the explosion. If they do, the outcome would be pretty much the same as in the Chandrasekhar mass models and GCE results would not change so much with respect 
to what we presented in this paper. A detailed analysis of double degenerate scenario as well as mergers as SN~Ia progenitors will be presented in a forthcoming paper.

Under the above conditions, we show here that SN~Ia can reproduce the
abundances of both $^{92}$Nb and $^{146}$Sm in meteorites within a factor of $\sim$2. The match would be poorer if solar system material had been isolated from fresh
nucleosynthetic inputs for a long time.

A detailed investigation of nuclear uncertainties affecting the reaction rates producing and destroying $^{92}$Nb, $^{92}$Mo, and $^{146}$Sm is presented. We found that the 
calculated $^{146}$Sm/$^{144}$Sm ratio was compatible with the meteoritic value when using a $^{148}$Gd($\gamma$,$\alpha$) rate based either on a fit to the Somorjai et al.~(1998) 
($\alpha$,$\gamma$) cross sections or on the recent rate including an additional reaction channel as presented by Rauscher (2013). Concerning $^{92}$Nb, the most important reactions 
affecting the $^{92}$Nb/$^{92}$Mo ratio were discussed and the impact of their nuclear uncertainties explored. The $^{92}$Nb/$^{92}$Mo ratio at 9.2 Gyr ranges between $1.66\times 
10^{-5}$ and $3.12\times 10^{-5}$ due to the nuclear uncertainties. This demonstrates that the meteoritic value can be reproduced within these uncertainties. We conclude that SNIa 
can play a key role in explaining meteoritic abundances of the extinct radioactivities $^{92}$Nb and $^{146}$Sm but that nuclear uncertainties still have considerable impact.

\acknowledgments We deeply thanks the anonymous referee for very usefull comments to this manuscript.
This work has been supported by the B2FH Association. The
numerical calculations have been supported by Regione Lombardia and CILEA
Consortium through a LISA Initiative (Laboratory for Interdisciplinary
Advanced Simulation) 2010 grant, and by R. Reifarth at Frankfurt Goethe
University. CT thanks C. Arlandini, P. Dagna, and R. Casalegno for technical
support in the simulations. TR is partially supported by the Swiss NSF, the
European Research Council, and the THEXO collaboration within the 7$^{th}$
Framework Programme of the EU. ND is supported by grants NNX12AH60G and NNX14AK09G from NASA. WH's work is supported by the German Science
Foundation (DFG) through the Cooperative Research Center TRR33 'Dark Universe'
and the Cluster of Excellence 'Origin and Structure of the Universe'. The work
of FR is supported by the Emmy Noether program of the German Science
Foundation (RO 3676/1-1), the ARCHES prize of the German Ministry of Education
and Research (BMBF) and by the Nuclear Astrophysics Virtual Institute
(VH-VI-417) of the Helmholtz Association.

\clearpage

\clearpage

\begin{deluxetable}{ccccccc}
\tabletypesize{\scriptsize}
\tablecaption{\label{tab:ratios}Radiogenic ratios at different metallicities}
\tablewidth{0pt}
\tablehead{
\colhead{Ratio} & \colhead{$Z$=0.003} & \colhead{$Z$=0.006} & \colhead{$Z$=0.01} &
\colhead{$Z$=0.012} & \colhead{$Z$=0.015} & \colhead{$Z$=0.02}}

\startdata

$^{92}$Nb/$^{92}$Mo & 7.363$\times$10$^{-4}$ & 1.145$\times$10$^{-3}$ & 1.526$\times$10$^{-3}$ & 1.322$\times$10$^{-3}$ & 1.846$\times$10$^{-3}$ & 1.635$\times$10$^{-3}$ \\
$^{97}$Tc/$^{98}$Ru & 1.215$\times$10$^{-2}$ & 1.767$\times$10$^{-2}$ & 2.354$\times$10$^{-2}$ & 2.406$\times$10$^{-2}$ & 2.533$\times$10$^{-2}$ & 2.285$\times$10$^{-2}$  \\
$^{98}$Tc/$^{98}$Ru & 8.465$\times$10$^{-5}$ & 1.798$\times$10$^{-4}$ & 3.384$\times$10$^{-4}$ & 3.711$\times$10$^{-4}$ & 4.741$\times$10$^{-4}$ & 5.066$\times$10$^{-4}$ \\
$^{146}$Sm/$^{144}$Sm & 4.053$\times$10$^{-1}$ & 3.705$\times$10$^{-1}$ & 3.624$\times$10$^{-1}$ & 3.762$\times$10$^{-1}$ & 3.329$\times$10$^{-1}$ & 3.161$\times$10$^{-1}$\\

\enddata

\vspace{0.1em}

\end{deluxetable}

\clearpage

\begin{deluxetable}{cccccc}
\tabletypesize{\scriptsize}
\rotate
\tablecaption{\label{tab:gce}Galactic chemical evolution of radiogenic isotopes}
\tablewidth{0pt}
\tablehead{
\colhead{[Fe/H]} & \colhead{Age (Gyr)} & \colhead{$^{92}$Nb/$^{92}$Mo} & \colhead{$^{97}$Tc/$^{98}$Ru} & \colhead{$^{98}$Tc/$^{98}$Ru} & \colhead{$^{146}$Sm/$^{144}$Sm } }

\startdata

 -1 & 1.57 & 1.700$\times$10$^{-4}$ & 1.752$\times$10$^{-5}$ & 2.582$\times$10$^{-6}$ & 4.209$\times$10$^{-2}$ \\
 -0.8 & 2.14 &  1.298$\times$10$^{-4}$ & 2.851$\times$10$^{-4}$ & 1.953$\times$10$^{-6}$ & 3.244$\times$10$^{-2}$ \\
 -0.5 & 3.50 & 8.344$\times$10$^{-5}$ & 1.778$\times$10$^{-4}$ & 1.557$\times$10$^{-6}$ & 2.039$\times$10$^{-2}$ \\
 -0.3 & 4.80 & 6.515$\times$10$^{-5}$ & 1.561$\times$10$^{-4}$ & 2.005$\times$10$^{-6}$ & 2.457$\times$10$^{-2}$ \\
 -0.155 & 6.20 & 4.033$\times$10$^{-5}$ & 1.018$\times$10$^{-4}$ & 1.432$\times$10$^{-6}$ & 1.807$\times$10$^{-2}$ \\
  0.0  & 9.20 & {\bf 1.752$\times$10$^{-5}$} & {\bf 4.077$\times$10$^{-5}$} & {\bf 6.471$\times$10$^{-7}$} & {\bf 6.989$\times$10$^{-3}$} \\
  0.092 & 11.7 & 1.137$\times$10$^{-5}$ & 2.681$\times$10$^{-5}$ & 4.803$\times$10$^{-7}$ & 4.644$\times$10$^{-3}$ \\
   & Meteoritic & (2.8$\pm$0.5)$\times$10$^{-5}$ &  $<$ 4.0$\times$10$^{-4}$ & $<$ 2.0$\times$10$^{-5}$ & (9.4$\pm$0.5)$\times$10$^{-3}$ \\

\enddata

\vspace{0.1em}





\vspace{0.1em}

\end{deluxetable}

\clearpage

\begin{deluxetable}{cccc}
\tabletypesize{\scriptsize}
\tablecaption{\label{tab:sm}Dependence of the $^{146}$Sm/$^{144}$Sm ratio on various $^{148}$Gd($\gamma$,$\alpha$) rates for SNIa at different metallicities and (last line) for GCE 
calculations.}
\tablewidth{0pt}
\tablehead{
\colhead{Z} & \colhead{RATH\tablenotemark{a}} & \colhead{exp ($\alpha$,$\gamma$) fit\tablenotemark{b}} & \colhead{2013\tablenotemark{c}} }

\startdata

0.003 & $4.053\times 10^{-1}$  &  $7.408\times 10^{-1}$ & $9.76\times10^{-1}$ \\
0.006 & $3.705\times10^{-1}$  &  $7.097\times 10^{-1}$ & $8.90\times10^{-1}$ \\
0.01 & $3.624\times10^{-1}$  &   $6.850\times 10^{-1}$ & $8.74\times10^{-1}$ \\
0.012 & $3.762\times10^{-1}$ &   $6.651\times 10^{-1}$ & $9.05\times10^{-1}$ \\
0.015 & $3.329\times10^{-1}$  &  $6.319\times 10^{-1}$ &  $8.01\times10^{-1}$ \\
0.02 & $3.161\times10^{-1}$ &    $6.132\times 10^{-1}$ & $7.62\times10^{-1}$ \\
\hline
GCE $\tau$=68 Myr& $6.989\times10^{-3}$ &  $1.050\times 10^{-2}$ &  $1.667\times10^{-2}$ \\

\enddata
\tablenotetext{a}{Rauscher \& Thielemann~(2000)}
\tablenotetext{b}{Somorjai et al.~(1998)}
\tablenotetext{c}{Rauscher~(2013)}
%
%
%

\end{deluxetable}

\clearpage

\begin{deluxetable}{ccc}
\tabletypesize{\scriptsize}
\tablecaption{\label{tab:nbmoratio}Reactions affecting the $^{92}$Nb/$^{92}$Mo ratio and their
variation to explore the nuclear uncertainties; rate set MIN yields the minimal ratio, set MAX the maximal ratio. The arrows indicate whether a rate has been multiplied by a factor of two
(arrow up) or divided by the same factor (arrow down). The modifications always apply to the rate and its reverse rate. In the last line there are the GCE calculations with these 
assumptions.}
\tablewidth{0pt} 
\tablehead{ \colhead{Reactions} & \colhead{Rate set MIN} & \colhead{Rate set MAX}}

\startdata
$^{91}$Zr(p,$\gamma$)$^{92}$Nb & $\downarrow$ & $\uparrow$ \\
$^{92}$Zr(p,$\gamma$)$^{93}$Nb & $\downarrow$ & $\uparrow$ \\
$^{92}$Zr(p,n)$^{92}$Nb & $\downarrow$ & $\uparrow$ \\
$^{91}$Nb(n,$\gamma$)$^{92}$Nb & $\uparrow$ & $\downarrow$  \\
$^{92}$Nb(n,$\gamma$)$^{93}$Nb & $\downarrow$ & $\uparrow$\\
&& \\
$^{91}$Nb(p,$\gamma$)$^{92}$Mo & $\uparrow$ & $\downarrow$  \\
$^{93}$Nb(p,n)$^{93}$Mo & $\uparrow$ & $\downarrow$  \\
$^{93}$Mo(n,$\gamma$)$^{94}$Mo & $\uparrow$ & $\downarrow$ \\
\hline
GCE & 1.660$\times$ 10$^{-5}$ & 3.118$\times$ 10$^{-5}$ \\

\enddata

\vspace{0.1em}

\end{deluxetable}

\end{document}